\documentclass[12pt,preprint,preprintnumbers,nofootinbib]{revtex4}

\usepackage{graphicx}
\usepackage{epsfig}    
\usepackage{dcolumn}
\usepackage{bm}
\usepackage{amssymb}
\usepackage{epsfig}    
\usepackage{here}

\newcommand{\nn}{\nonumber}

\newcommand{\al}{\alpha}
\newcommand{\be}{\beta}

\newcommand{\gsim}{\mbox{ \raisebox{-1.0ex}{$\stackrel{\textstyle >}
{\textstyle \sim}$ }}}
\newcommand{\lsim}{\mbox{ \raisebox{-1.0ex}{$\stackrel{\textstyle <}
{\textstyle \sim}$ }}}

\def\Journal#1#2#3#4{{#1} {\bf #2} (#4) #3}

\def\NPB{{\em Nucl. Phys.} B}
\def\PLB{{\em Phys. Lett.}  B}
\def\PRL{\em Phys. Rev. Lett.}
 
\def\PRD{{\em Phys. Rev.} D}

\def\EPC{{\em Euro. Phys. J.} C}

\begin{document}
\topmargin -1cm

\begin{flushright}
KEK-TH-856\\
MSUHEP-21119 
\end{flushright}

\vspace*{4mm}

\begin{center}
{\large \bf New Physics Effect on the Higgs Self-Coupling}

\vspace*{4mm}
{{\sc Shinya Kanemura}$^{a,\,}$\footnote{
                   E-mail: shinya.kanemura@kek.jp}, 
        {\sc Shingo Kiyoura}$^{a,b,\,}$\footnote{
                   E-mail: shingo.kiyoura@kek.jp},  
        {\sc Yasuhiro Okada}$^{a,c,\,}$\footnote{
                   E-mail: yasuhiro.okada@kek.jp},  \\
        {\sc Eibun Senaha}$^{a,c,\,}$\footnote{E-mail: senaha@post.kek.jp}, 
        {\sc C.-P. Yuan}$^{d,\,}$\footnote{E-mail: yuan@pa.msu.edu}}    

\vspace*{4mm}
{
{\em $^a$Theory Group, KEK, Tsukuba, Ibaraki 305-0801, Japan\\
     $^b$Department of Radiological Science, Ibaraki Prefectural University 
         of Health Sciences, Ami, Inashiki, Ibaraki 300-0394, Japan\\
     $^c$Department of Particle and Nuclear Physics, 
         the Graduate University for Advanced Studies, Tsukuba, 
         Ibaraki 305-0801, Japan\\ 
     $^d$Physics and Astronomy Department, Michigan State University, 
     East Lansing, Michigan 48824-1116, USA}}
\end{center}

\begin{abstract}
One-loop corrections to the self-coupling constant of the lightest 
CP-even Higgs boson are studied in the two Higgs doublet model. 
After renormalization is performed, quartic contributions 
of heavy particle's mass can appear in the effective coupling. 
We find that these non-decoupling effects can yield ${\cal O}(100)$ 
\% deviations from the Standard Model prediction, even when all the 
other couplings of the lightest Higgs boson to gauge bosons and 
fermions are in good agreement with the Standard Model. 
\end{abstract}

\maketitle


\section{Introduction}

The cause of the electroweak symmetry breaking 
(that generates masses for the weak gauge bosons 
$W^{\pm }$ and $Z$) 
and the origin of the flavor symmetry 
breaking (that generates mass spectrum for quarks and leptons) 
are great mysteries in the elementary particle physics. 
In the Standard Model (SM), both symmetry breaking mechanisms 
are explained by introducing  
a scalar iso-doublet field which includes a physical scalar state, 
i.e., the Higgs boson ($h$).  
The $W$ and $Z$ bosons obtain their masses through the Higgs mechanism,  
and the fermions gain their masses via the Yukawa interaction. 

The present precision data have shown an excellent agreement with the
prediction of the SM \cite{LEPEWWG,QCD}. 
It also suggests the Higgs boson to be light, with a 
central value $m_H=81$\,GeV (below the LEP2 direct search
bound which requires $m_H > 114.4$\,GeV \cite{direct_search}) 
and a 95\%\,C.L. upper limit on Higgs boson mass to be 
193 GeV \cite{EWWG}.
However, a more close examination reveals that the present data only
strongly constrains the couplings of the gauge bosons with the
fermions (except top quark) as well as 
those of the triple gauge boson vertices, 
but says little about the interaction of
Higgs boson with gauge bosons and fermions. 
With the data coming from the current Run-II of the Fermilab 
Tevatron collider, the CERN Large Hadron Collider (LHC), 
and future Linear Colliders (LC's), more precise tests of these sectors
of the SM become possible. 
For example, at the LC, the Higgs boson can be produced
via the processes $e^+e^- \to Z^\ast \to Z h$ 
and 
$e^+e^- \to {W^+}^\ast \bar{\nu} {W^-}^\ast \nu \to
h \bar{\nu}\nu$ \cite{eehx}, and 
the $hZZ$ and $hW^+W^-$ couplings can be determined 
to a few percent 
from the precise measurement of the above production 
cross sections \cite{LC}. 
Furthermore, the Yukawa couplings of the Higgs boson to fermions can be 
determined from measuring the decay branching ratios of the Higgs boson.

To test the electroweak symmetry breaking (EWSB) sector of the SM,
not only should the couplings of the Higgs boson to gauge 
bosons and fermions be measured, but also the self-coupling of the Higgs
boson which originates from the Higgs potential.
Unfortunately, at the LHC, it would be extremely difficult to measure 
the SM Higgs boson self-couplings, either the trilinear coupling    
$hhh$ or the quartic coupling $hhhh$ \cite{hhh_LHC}. 
At the LC, the trilinear coupling $\lambda_{hhh}^{}$ can be 
measured via the Higgs boson pair production in 
$e^+e^- \to Z^\ast \to Z hh$ and
$e^+e^- \to {W^+}^\ast \bar{\nu} {W^-}^\ast \nu \to hh \bar{\nu}\nu$,
if the Higgs boson is not too heavy \cite{eehhx,eehhx2,battaglia,hhh_ACFA}.
It is expected that at a $500$\,GeV ($3$\,TeV) $e^+e^-$ collider  
with an integrated luminosity of $1$ ab$^{-1}$
($5$ ab$^{-1}$), $\lambda_{hhh}^{}$ can be measured to about 20\% (7\%)
accuracy for the Higgs boson mass around $120$ GeV \cite{battaglia}. 
The expectation for such a precise measurement motivates to study 
radiative correction to the Higgs self-couplings. 
In the SM, the one-loop contribution of the heavy quarks
is substantial due to their non-decoupling property, especially its 
leading effect grows as the quartic power of the top-quark mass 
in the large mass limit.   
The precise measurement of the self-coupling at the LC also makes 
it possible to test extended Higgs models, which have different 
structures of the Higgs potential from the 
SM \cite{eehhx2,hhh_ext}.  

In this Letter, we discuss quantum corrections to the 
self-coupling of the lightest CP-even Higgs boson in the 
two-Higgs-doublet model (THDM) \cite{lcws02kanemu}. 
The THDM is the simplest model of extended Higgs sectors, and its Higgs 
potential has a rich structure for various physics motivations. 
The Higgs sector of the Minimal Supersymmetric Standard Model (MSSM) 
is a special case of the {\it weakly coupled} THDM \cite{HHG}.
Some models of dynamical electroweak symmetry breaking 
also yield the THDM as their low-energy effective theory \cite{TC}, 
in which the Higgs self-couplings are relatively strong. 
The tree-level $hhh$ coupling in the THDM generally differs from  
that predicted by the SM, 
depending on the other parameters of the model \cite{eehhx2,hhh_ext}.  
In Ref.~\cite{hhh_hollik}, 
the one-loop effects on the Higgs self-couplings in the MSSM   
have been calculated, and their decoupling property
has also been studied in detail. 
As to be shown below, in contrast to the MSSM, it can happen 
that the heavy mass effects in a general THDM do not decouple.
To illustrate this point, 
we calculate the Higgs boson self-coupling by the diagrammatic approach as well 
as the effective potential method.
After rewriting the one-loop effect in terms of the 
renormalized mass of the lightest Higgs boson, 
quartic dependence on the mass of the heavier Higgs bosons 
contributing in the loop diagrams appears in the effective $hhh$ coupling. 
We find that even when the Higgs couplings to gauge bosons and fermions 
are almost SM-like, the deviation in the $hhh$ self-coupling 
from the SM prediction can be at the order of $100\%$ due to    
the non-decoupling effects of the additional heavier Higgs bosons in loops.

\section{Leading SM corrections to the $hhh$ coupling}

Before proceeding to the discussion of the results for the THDM, 
it would be instructive 
to see how the leading one-loop effects appear in the $hhh$ coupling 
in the SM. The tree-level trilinear Higgs coupling is expressed in terms of 
the Higgs boson mass ($m_h^{}$) and the vacuum expectation value ($v$) by 
$\lambda_{hhh}^{tree}(SM) = \frac{3 m_h^2}{v}$. 
The leading one-loop contribution of the top quarks to the 
effective coupling $\lambda_{hhh}^{eff}(SM)$ is derived as 
\begin{eqnarray}
  \lambda_{hhh}^{eff}(SM) =  \frac{3 m_h^2}{v}
       \left[\, 1 - \frac{N_{c}}{3 \pi^2} \frac{m_t^4}{v^2 m_h^2}
     \left\{ 1 + {\cal O}\left(\frac{m_h^2}{m_t^2},\; 
                              \frac{p_i^2}{m_t^2} \right) \right\} 
\,\right], \label{smhhh}
\label{mt4term_hhh_SM}
\end{eqnarray}
where $m_h^{}$ and $m_t^{}$ are the physical masses of the Higgs boson 
and the top-quark, respectively,  
and $p_i^{}$ ($i=1$-$3$) represent the momenta of external Higgs lines.
As an interesting feature, the non-vanishing one-loop effect 
of the top-quarks appears as ${\cal O}(m_t^4)$.    

The easiest way to understand the appearance of the quartic mass 
term in the effective $hhh$ coupling is to study  
the one-loop effective potential; 
$V_{eff}[\varphi] = V_{\rm tree}[\varphi] + \Delta V[\varphi]$. 
The one-loop contribution $\Delta V[\varphi]$ is given by 
\begin{eqnarray}
   \Delta V[\varphi] = \frac{1}{64\pi^2} \sum_f 
                     N_{c_f} N_{s_f} (-1)^{2 s_f}  
                     (M_f[\varphi])^4 
             \left\{ \ln \frac{(M_f[\varphi])^2}{Q^2}
                         -\frac{3}{2}\right\},  
\label{eff}
\end{eqnarray}
where $\varphi=\langle \phi \rangle = v + \langle h \rangle$, 
$N_{c_f}$ is the color number,   
$s_f$ ($N_{s_f}$) is the spin (degree of freedom) of the field $f$  
in the loop, $M_f[\varphi]$ is the field dependent mass of $f$, 
and $Q$ is an arbitrary energy scale.  The effective coupling of 
$hhh$ can be expressed in terms of the physical mass of the Higgs 
boson and $\Delta V[\varphi]$ by  
\begin{eqnarray}
\lambda^{eff}_{hhh} = 
\left.\frac{\partial^3 V_{eff}}{\partial \varphi^3}\right|_v = 
\left. \frac{3 m_h^2}{v}   +   
 \left(  \frac{3}{v^2} \frac{\partial}  {\partial \varphi} 
       - \frac{3}{v}   \frac{\partial^2}{\partial \varphi^2} 
       + \frac{\partial^3}{\partial \varphi^3} 
 \right) \Delta V[\varphi]\right|_v, \label{dec_hhh} 
\end{eqnarray}
up to the wave function renormalization contributions.
We note that $m_h^2$ is obtained as the second derivative of 
$V_{eff}^{}[\varphi]$. 
The leading top-quark loop effect in Eq.~(\ref{smhhh}) is 
easily obtained from Eq.~(\ref{dec_hhh}) 
with the field dependent mass $M_t[\varphi] = y_t \frac{\varphi}{\sqrt{2}}$,   
where $y_t^{}$ is the top-Yukawa coupling constant in the SM. 
Although each individual term inside the parenthesis (present in 
the right-hand side) of Eq.~(\ref{dec_hhh}) can contribute a large 
logarithmic term  $m_t^4\ln(m_t^2/Q^2)$, they all cancel with each other 
in the sum so that the remaining leading contribution to 
$\lambda^{eff}_{hhh}$ is the constant $m_t^4$ term. 

The appearance of this non-vanishing $m_t^4$ term is a striking 
feature of the one-loop correction to the self-coupling constant.  
In contrast, the one-loop effective couplings of $hVV$ ($VV=ZZ$, $W^+W^-$)  
($g_{hVV}^{}$) have non-decoupling power-like contributions of 
at most ${\cal O}(m_t^2)$.
Therefore, 
if a new heavy particle has the similar non-decoupling property to 
the top quark in some extension of the SM, 
its loop effect on the $hhh$ coupling can become important, 
because the quartic mass contribution is expected to make the 
correction large.
In the following, we examine this point in the context of the THDM.

\section{One loop effect on the $hhh$ coupling in the THDM}

\indent
The Higgs potential of the CP-conserving THDM is given by
\begin{eqnarray}
  {V}_{\rm THDM}  &=&     m_1^2 \left| \Phi_1 \right|^2 
                          + m_2^2 \left| \Phi_2 \right|^2 - 
                              m_3^2 \left( \Phi_1^{\dagger} \Phi_2 
                                + \Phi_2^{\dagger} \Phi_1 \right) 
                                \nn  
                       + \frac{\lambda_1}{2} 
                               \left| \Phi_1 \right|^4 
                             + \frac{\lambda_2}{2} 
                               \left| \Phi_2 \right|^4 \nn \\
& &                          + \lambda_3 \left| \Phi_1 \right|^2 
                                \left| \Phi_2 \right|^2 
                      + \lambda_4 
                               \left| \Phi_1^{\dagger} \Phi_2 \right|^2
                             + \frac{\lambda_5}{2} 
                             \left\{ 
                               \left( \Phi_1^{\dagger} \Phi_2 \right)^2
                            +  \left( \Phi_2^{\dagger} \Phi_1 \right)^2
                             \right\},  \label{pot}
\end{eqnarray}
where we imposed a softly-broken discrete symmetry under the 
transformation of $\Phi_1 \to \Phi_1$, $\Phi_2 \to - \Phi_2$. 
We assume all the coupling constants and  
mass parameters to be real, so that there are eight real 
parameters in the potential (\ref{pot}).
The discrete symmetry allows two types of the Yukawa interaction, 
so called Model I and Model II \cite{HHG}.
The discrete symmetry ensures natural suppression of 
flavor changing neutral current processes. 

Diagonalizing the mass matrices, we have
the five physical scalar states; i.e., 
two CP-even ($h$, $H$), one CP-odd ($A$),  
and a pair of charged  ($H^\pm$) Higgs bosons.  
Notice that masses of the heavier Higgs bosons 
($H$, $H^\pm$ and $A$) schematically take the form as 
\begin{eqnarray}
  m_{\Phi}^2 \simeq M^2 + \lambda_i v^2,   \label{typ_mass}
\end{eqnarray}
where   
$\Phi$ represents $H$, $H^\pm$ or $A$, 
$M$ is the soft-breaking scale of the discrete symmetry  
defined by $M=m_3/\sqrt{\sin\be\cos\be}$, and 
$\lambda_i$ is some linear combination of 
$\lambda_1$-$\lambda_5$.
As indicated in Eq.~(\ref{typ_mass}), 
there are two origins of the masses; 
one is the soft-breaking scale $M$, and  
another is the vacuum expectation value 
of the electroweak symmetry breaking $v$.
The origin of the mass determines the decoupling
property of the heavy Higgs bosons \cite{nondec3}.
When $M^2 \gg \lambda_i v^2$, 
$m_{\Phi}^{}$ is determined by $M$, and is independent of $\lambda_i$. 
Consequently, the loop effects of $\Phi$ vanish in the large mass limit 
(i.e., $m_{\Phi}^{} \to \infty$) because of the decoupling 
theorem\footnote{The MSSM Higgs sector corresponds to this case, 
in which coupling constants $\lambda_1$-$\lambda_5$ are given to be 
${\cal O}(g^2_i)$ due to supersymmetry, 
where $g_i$ represent the electroweak gauge coupling constants.}. 
On the other hand, when $M^2 \lsim \lambda_i v^2$ the large value  
of $m_\Phi^{}$ is realized by large coupling constants $\lambda_i$.  
In this case, the decoupling theorem cannot be applied.
Hence, we expect positive power (or logarithmic)  
contributions of $m_{\Phi}^{}$ in the radiative 
correction \cite{nondec_2HDM,SMlikeTHDM}. 
We refer such power-like contribution as the non-decoupling effect.
In these scenarios, theoretical consistencies and present experimental 
data generally provide strong constraints to the model parameters. 
For instance, too large values of $\lambda_i$ clearly break validity 
of perturbation calculation. 

Now we discuss the $hhh$ coupling in the THDM. 
At the tree level, 
the $hhh$ coupling is expressed in terms of the input parameters 
of the Higgs sector by 
\begin{eqnarray}
\lambda_{hhh}^{tree}(THDM)&=& -\frac{3}{4 v \cos\be\sin\be} 
     \left[ 4 M^2 \cos^2(\al-\be) \cos(\al+\be)\right.\nonumber\\
&&\left.  - \left\{ \cos(3\al-\be)+ 3 \cos(\al+\be) \right\} m_h^2 \right].
\label{treehhh}
\end{eqnarray}
For general values of $\alpha-\beta$, the coupling depends on 
the mixing angles and the soft-breaking scale of the discrete symmetry, 
and thus its value can be completely different from the SM prediction. 
In general, when $\lambda_{hhh}^{tree}(THDM)$ is 
significantly different from $\lambda_{hhh}^{tree}(SM)$, the 
new physics effect will also manifest in the coupling of 
Higgs bosons to gauge bosons and fermions, which can be detected
experimentally at future high energy colliders. 

We have calculated the one-loop correction to the effective $hhh$ 
vertex function by the diagrammatic approach in the on-shell scheme.  
In this Letter, we present the results of our calculations, and  
details of the calculation will be shown elsewhere \cite{fullpaper}.
To simplify our discussion on the 
one-loop radiative corrections to the Higgs self-coupling 
in the THDM, we shall assume the following scenario:
({\it i}) Only one Higgs boson ($h$) is found and 
    its mass ($m_h$) is measured to be light ($\lsim 200$ GeV). 
({\it ii}) Experimental data on the $hVV$ couplings ($g_{hVV}^{}$) 
    and the the Higgs 
    decay branching ratios agree with their SM prediction 
    in good accuracy. 
This implies $\sin^2(\alpha-\beta) \simeq 1$ 
(or $\alpha\simeq\beta-\pi/2$) in the context of 
the THDM \cite{SMlikeTHDM}.
In this case, the tree $hhh$ coupling given in 
Eq.~(\ref{treehhh}) takes the same form as the SM prediction; i.e., 
$\lambda_{hhh}^{tree}(THDM) = \frac{3 m_h^2}{v}$.

Using the Feynman diagrammatic method, we calculate 
the leading contributions originated from the heavy Higgs 
boson loops and the top quark loops.
We find that at the one loop level, the effective $hhh$ 
coupling can be written as    
\begin{eqnarray}
 \lambda_{hhh}^{eff}(THDM) \!\!&=&\!\! \frac{3 m_h^2}{v}
      \left\{ 1  
              + \frac{m_{H}^4}{12 \pi^2 m_h^2 v^2} 
                         \left(1 - \frac{M^2}{m_H^2}\right)^3 
              + \frac{m_{A}^4}{12 \pi^2 m_h^2 v^2} 
                         \left(1 - \frac{M^2}{m_A^2}\right)^3 \right.\nn\\
&&\left.
\!\!\!\!\!\!\!\!\!\!\!\!\!\!\!\!\!\!\!\!\!\!\!\!\!\!\!\!\!\!
\!\!\!\!\!\!\!\!\!\!\!\!\!\!\!\!\!\!
              + \frac{m_{H^\pm}^4}{6 \pi^2 m_h^2 v^2} 
                         \left(1 - \frac{M^2}{m_{H^\pm}^2}\right)^3
              - \frac{N_{c_t} m_t^4}{3 \pi^2 m_h^2 v^2} + 
              {\cal O} \left(\frac{p^2_i m_\Phi^2}{m_h^2 v^2},
                           \;\frac{m_\Phi^2}{v^2},
                           \;\frac{p^2_i m_t^2}{m_h^2 v^2},  
                           \;\frac{m_t^2}{v^2}  \right)
      \right\}, \label{m4THDM}    
\end{eqnarray}
where 
$m_\Phi^{}$ and $p_i$ represent the mass of $H$, $A$ or $H^\pm$ 
and the momenta of external Higgs lines, respectively.     
We note that in Eq.~(\ref{m4THDM}) $m_h$ is the renormalized 
physical mass of the lightest CP-even Higgs boson $h$. 
The leading contribution of the above result can also 
be obtained by using the effective potential method. 
As expected, the contribution from the top quark loop is the same 
as the SM prediction because the tree level coupling of the top 
quark to $h$ is identical to that in the SM when 
$\alpha=\beta-\pi/2$. Similar to the top quark loop, the 
contribution from the Higgs boson loops also grows as $m_{\Phi}^4$ 
but with a suppression factor $(1-M^2/m_{\Phi}^2)^3$, where $\Phi$ 
represents $H$, $A$ or $H^\pm$. Furthermore, the Higgs boson loop 
contributes an opposite sign to the top quark loop because 
the former is a boson loop and the latter is a fermion loop.
Because of the suppression factor $(1-M^2/m_{\Phi}^2)^3$, 
the maximum non-decoupling 
effect is realized in the limit of $M^2 \to 0$, and the 
Higgs boson loop contribution is enhanced by $m_{\Phi}^4$.
On the other hand, when the Higgs boson mass $m_{\Phi}$
is at the same order as the soft mass scale $M$, the 
Higgs boson loop contribution becomes diminished and
decoupled from $\lambda_{hhh}^{eff}(THDM)$.

To examine the numerical effect of the one loop radiative corrections
to the trilinear coupling $\lambda_{hhh}$ in the THDM, we have to     
take into account various theoretical and experimental constraints.
Some of them are discussed below.
The choice of the THDM parameters should satisfy the requirement of
perturbative unitarity for the $S$-wave amplitudes
of the $2 \to 2$ scattering processes
of the Higgs bosons and longitudinally
polarized weak bosons \cite{LQT}. For example, the unitarity condition
requires that when $m_h=120$ GeV, $m_A^{}=m_H^{}=m_{H^\pm}\;
(\equiv m_\Phi^{})$ and $M=0$, the upper bound on the
masses of the heavier Higgs bosons is about $550$\,GeV to
$600$\,GeV~\cite{unitarity}.
This upper bound is generally weakened
as $M$ increases, so a heavier Higgs boson is allowed. 
We also include the constraints imposed from the low-energy
precision data on the THDM \cite{ST_2HDM}, especially,
the $\rho$ parameter constraint ($\Delta \rho (\equiv \rho-1) \sim 10^{-3}$).
To satisfy this constraint, the THDM has to have an
approximate custodial ($SU(2)_V^{}$) symmetry \cite{CSTHDM}.
In the Higgs sector of the THDM, 
there are typically two options of the parameter 
choice in which $SU(2)_V^{}$ is conserved according to the 
assignment of the $SU(2)_V$ charge; 
(1) $m_{H^\pm}^{} \simeq m_A^{}$, and 
(2) $m_{H^\pm}^{} \simeq m_H^{}$ with $\sin^2(\alpha-\beta) \simeq 1$ 
 or $m_{H^\pm}^{} \simeq m_h^{}$ with 
$\cos^2(\alpha-\beta) \simeq 1$ \cite{CSTHDM,rhoTHDM}\footnote{
In terms of the coupling constants, these conditions are 
expressed by (1) $\lambda_4=\lambda_5$, and (2) 
$\lambda_1=\lambda_2=\lambda_3$ with $m_1^2=m_2^2$.}. 
In our numerical analysis, we choose the parameters that satisfy 
these conditions.
In addition, in Model II of the THDM, it is known that the 
$b \to s \gamma$ branching ratio excludes the small mass 
of the charged Higgs boson \cite{bsa-2hdm}. 

\begin{figure}[t]
\vspace*{-12mm}
\begin{center}
\hspace*{-3mm}
\includegraphics[width=10cm,height=8cm]{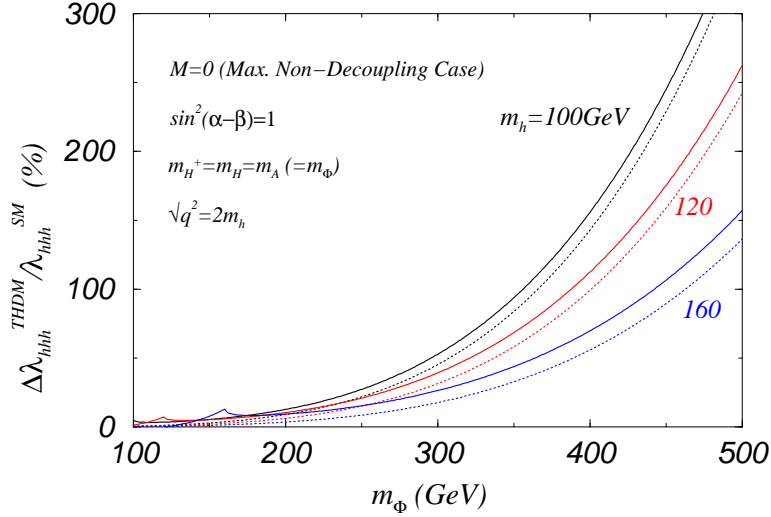}
\end{center}
\vspace*{-8mm}
\caption{The $m_\Phi^4$ behavior in  
              $\Delta\lambda_{hhh}^{THDM}$, 
              where  $\Delta\lambda_{hhh}^{THDM}$ 
              $\equiv \lambda_{hhh}^{eff}(THDM)-\lambda_{hhh}^{eff}(SM)$.  
              The results of the full one-loop calculation 
              are shown as solid curves, while the quartic mass 
              contributions given in 
              Eq.~(\ref{m4THDM}) are plotted as dotted curves.}
\end{figure}

In Fig.~1, we show $\Delta\lambda_{hhh}^{THDM}/ \lambda_{hhh}^{eff}(SM)$ as
a function of $m_\Phi^{}$ (=$m_H^{}=m_A^{}=m_{H^\pm}^{}$)
in the SM-like scenario ($\sin^2(\al-\be)=1$) 
for $m_h=100$, $120$ and $160$ GeV, where
$\Delta\lambda_{hhh}^{THDM} 
\equiv \lambda_{hhh}^{eff}(THDM)-\lambda_{hhh}^{eff}(SM)$.
The solid curves are the results from a full calculation of the
Higgs boson loop contributions, and the dotted curves are the
leading contributions in Eq.~(\ref{m4THDM}). 
Here, we have chosen $M=0$ to explore the maximal non-decoupling effect.
Because of the quartic power dependence of $m_\Phi^{}$, the
non-decoupling effect becomes greater
for larger values of $m_\Phi^{}$ with smaller $m_h$.
However, the allowed value of $m_\Phi^{}$ 
is bounded
 from above by the perturbative unitarity bound
($m_\Phi^{} \lsim 550$-$600$ GeV in this case). 
As shown in Fig.~1, the deviation from the SM prediction is
about 30\% (100\%) for $m_\Phi^{}=300$ (400) GeV, in the maximal
non-decoupling scenario\footnote{
Although the expression in Eq.~(\ref{m4THDM}) does not depend on
$\tan\beta$, the allowed value of $\tan\beta$ is constrained to be
${\cal O}(1)$ due to the requirement of 
the perturbative unitarity when large values of $m_\Phi^{}$
are taken with $M=0$.
Hence, the large deviation from the SM prediction occurs  
at $\tan\beta = {\cal O}(1)$. 
We note that the parameter set 
$m_H^{}=m_A^{}=m_{H^\pm}^{}$, $M=0$, $\alpha=\beta-\pi/2$ and 
$\tan\beta=1$ 
corresponds to $\lambda_1=\lambda_2=\lambda_3=(m_h^2+m_H^2)/v^2$ and 
$\lambda_4=\lambda_5=-m_H^2/v^2$.  
}.
We note that the large, of ${\cal O}(1)$, one-loop radiative correction to
$\lambda_{hhh}$ in the THDM does not imply the breakdown
of the perturbative expansion, for the large contribution originates
from new types of couplings,
e.g., $\lambda_{h\Phi\Phi}^{}$ and $\lambda_{hh\Phi\Phi}$,
 that enter in loop calculations. Needless to say that we do not expect
such kind of large correction to occur beyond the one-loop order. 

\begin{figure}[t]
\vspace*{-12mm}
\begin{center}
\hspace*{-3mm}
\includegraphics[width=10cm,height=8cm]{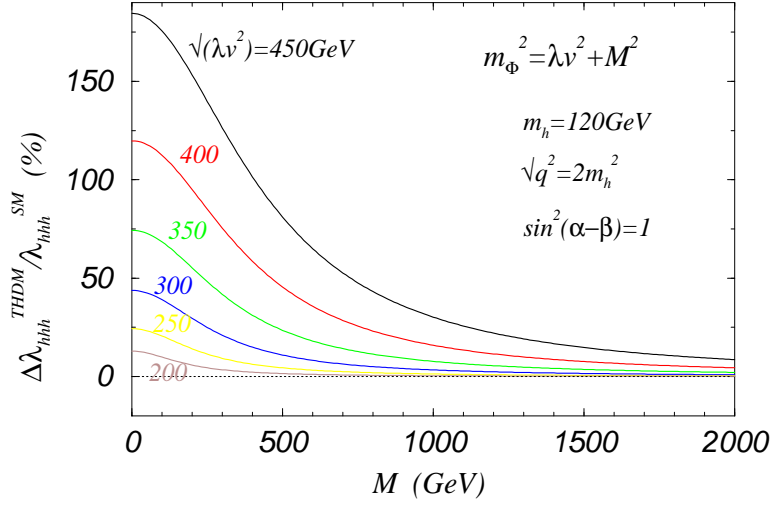}
\end{center}
\vspace*{-8mm}
\caption{     The decoupling behavior of   
              $\Delta\lambda_{hhh}^{THDM}$, 
              where $\Delta\lambda_{hhh}^{THDM}$ is defined by 
              $\lambda_{hhh}^{eff}(THDM)-\lambda_{hhh}^{eff}(SM)$ 
              calculated in the one-loop diagrammatic approach. 
              The mass of the heavy Higgs bosons 
              $m_{\Phi} (\equiv m_H^{}=m_A^{}=m_{H^\pm}^{})$ 
              is given by $m_{\Phi}^2=\lambda v^2+M^2$.         }
\end{figure}

As shown in Eq.~(\ref{m4THDM}),
the non-decoupling effect of the heavier Higgs bosons is suppressed
by the factor of $(1-\frac{M^2}{m_\Phi^2})^3$ for a non-vanishing $M$.
In the case of $M^2 \gg \lambda_i v^2$, 
cf. Eq~(\ref{typ_mass}), this factor behaves as  
\begin{eqnarray}
  \frac{1}{16\pi^2} \frac{m_\Phi^4}{v^2 m_h^2}
   \left( 1- \frac{M^2}{m_\Phi^2}\right)^3
 \longrightarrow
   \frac{\lambda_i^3}{16\pi^2} \frac{v^2}{m_h^2} \frac{v^2}{m_\Phi^2},
\end{eqnarray}
and 
thus decouples in the limit of $m_\Phi^2 (\simeq M^2) \to \infty$.
In Fig.~2, we show the decoupling behavior of the heavier Higgs 
contribution
as a function of $M$ with fixed $\sqrt{\lambda v^2}=200-450$ GeV, in the
case of $\sin^2(\al-\be)=1$ and $m_h^{}=120$ GeV, where
the mass of the heavier Higgs bosons
$m_{\Phi}^{}$ ($=m_A^{}=m_H^{}=m_{H^\pm}$)
is given by $m_\Phi^2=\lambda v^2 + M^2$. 
(We note that $\lambda$ corresponds to 
 $\lambda_1 \cos^2 \beta + \lambda_2 \sin^2\beta - m_h^2/v^2 
  = \lambda_3 - m_h^2/v^2 = 
  - \lambda_4 = - \lambda_5$ in this case.)
It is evident that
the heavier Higgs boson contributions reduce rapidly for a larger
value of $M$. Nevertheless, 
a few tens of percent of the correction remains at
$M=1000$ GeV.
Since the Higgs sector of the MSSM is a special case of the
Type-II THDM with $\lambda_i v^2 \simeq {\cal O}(m_W^2)$,
as required by supersymmetry, it belongs to the class of
models in which the heavier Higgs bosons decouple.
Hence, the effect of the Higgs boson loops to
$\lambda_{hhh}^{eff}(MSSM)$ is expected to be small.  
A detailed study on this decoupling
behavior of the one-loop corrected $hhh$ coupling
in the MSSM can be found in Ref.~\cite{hhh_hollik}. 
We confirmed that our results for large values of $M$ 
are consistent with 
those in Ref.~\cite{hhh_hollik}. 

\begin{figure}[t]
\vspace*{-12mm}
\begin{center}
\hspace*{-3mm}
\includegraphics[width=10cm,height=8cm]{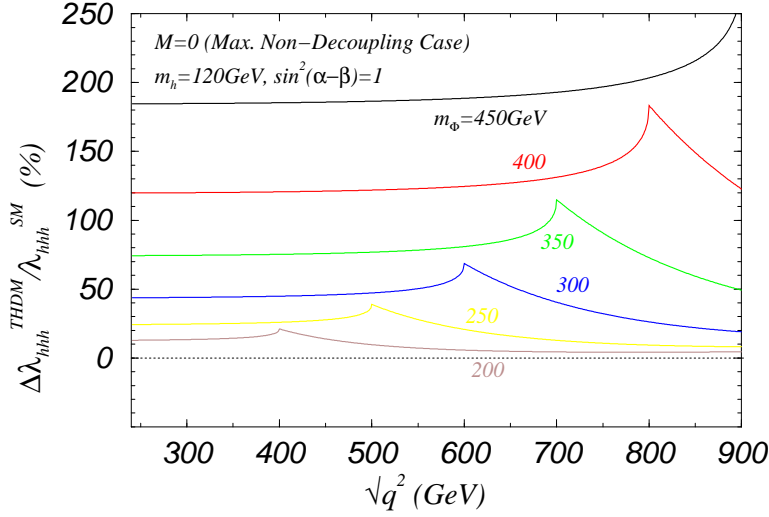}
\end{center}
\vspace*{-8mm}
\caption{     The momentum dependence of 
          $\Delta\lambda_{hhh}^{THDM}(q^2) \equiv
           \lambda_{hhh}^{eff}(THDM)- \lambda_{hhh}^{eff}(SM)$ 
          calculated in the diagrammatic approach, where  
          $\sqrt{q^2}$ is the invariant mass of $h^\ast$ in 
          $h^\ast \to hh$ for each value
          of $m_\Phi^{}$ ($\equiv m_H^{}=m_A^{}=m_{H^\pm}^{}$), 
          when $m_h=120$ GeV, $\sin(\alpha-\beta)=-1$ and $M=0$.}
\end{figure}

New physics model can strongly modify the
trilinear coupling of the Higgs boson, which can be
measured from studying the scattering processes 
$e^+e^- \to Z^\ast \to Z h^\ast \to Z hh$
and $e^+e^- \to \bar \nu \nu W^{+\ast} W^{-\ast}
            \to \bar \nu \nu  h^\ast
            \to \bar \nu \nu  h  h$ in the $e^+e^-$ collision,
and $\gamma\gamma \to h^\ast \to hh$
at the $\gamma\gamma$ option of the LC.~In 
Fig.~3, we show the momentum dependent effective self-coupling
of the Higgs boson, 
$\lambda_{hhh}^{eff}(q^2)$, as a function of the invariant mass
($\sqrt{q^2}$) of the virtual $h$ boson, for
various values of $m_\Phi^{}$ ($=m_A^{}=m_H^{}=m_{H^\pm}$)
with $\sin^2(\al-\be)=1$ and $m_h=120$ GeV.
Again, to show the maximal non-decoupling effect,
we have set $M$ to be zero.
The Higgs boson one-loop contribution is always positive. 
Below the peak of the threshold of the heavy Higgs pair production, 
$\lambda_{hhh}^{eff}(q^2)$ is insensitive to $\sqrt{q^2}$. 
We note that the low $\sqrt{q^2}$  (but $\sqrt{q^2} \gsim 2 m_h$) 
is the most important region in the extraction of the $hhh$ 
coupling from the data of the double Higgs production mechanism, 
because the $h^\ast$ propagator $1/(q^2-m_h^2)$ in the signal 
process becomes larger.  
On the contrary, the fermionic (top-quark) loop effect 
strongly depends on $\sqrt{q^2}$, because the threshold enhancement 
at $\sqrt{q^2}=2 m_t^{}$ contributes an opposite sign 
to the quartic mass term contribution.  
Including both the scalar and the fermion loop contributions, 
the one-loop radiative correction to the $hhh$ coupling 
changes sign when $\sqrt{q^2}$ is somewhere 
between $2 m_h$ and $2 m_t$.  

Finally, we comment on the cases in which the assumption 
$\sin^2(\alpha-\beta)=1$ is slightly relaxed; i.e.,  
the coupling for ${hVV}^{}$ ($VV=ZZ$ and $W^+W^-$) deviates from 
the SM prediction by a factor of $\sin(\beta-\alpha)$ at the tree level. 
When $g_{hVV}^{}$  are measured to be nearly 
(not exactly) SM-like with a few percent deviation, the {\it tree}-level 
$hhh$ coupling can also be different from the SM prediction for the 
given $m_h^{}$, depending on the soft-breaking scale $M$ 
for the discrete symmetry as well as $\tan\beta$, 
as shown in Eq.~(\ref{treehhh}).   
By scanning the parameters $m_h^{}$ and $\sin(\alpha-\beta)$ 
under the available phenomenological 
constraints and the theoretical bounds, such as the perturbative 
unitarity and vacuum stability, 
one can obtain the range of the allowed deviation 
in the tree-level $hhh$ coupling from the SM prediction. 
When the measured $hhh$ coupling is found to be out of the 
allowed range of the tree-level deviation for each set of the parameters 
$m_h^{}$ and $\sin(\alpha-\beta)$, 
the one-loop effect to the $hhh$ coupling can be readily
identified.
Especially, in the case of $M^2 \lsim \lambda_i v^2$, we found that 
the tree-level $hhh$ coupling can at most differ 
from that in the SM by about 10\%, 
assuming $g_{hVV}^{}$ only deviates from the SM prediction by a few percent.   
Hence, a large positive deviation arising from the one-loop 
${\cal O}(m_A^4)$ contribution can be much larger than the 
tree-level deviation from the SM $hhh$ coupling, as $\sin^2(\alpha-\beta)$ 
slightly deviates from 1.
A more detailed discussion on this point will be presented 
in Ref.~\cite{fullpaper}.

\section{Conclusion}

\indent
We have examined the one-loop correction to the Higgs 
self-coupling $hhh$ in the THDM. There can be non-decoupling 
quartic mass contributions of the heavier Higgs bosons 
in the $hhh$ coupling. 
Because of these effects, deviation from the SM 
prediction can be ${\cal O}(100)$\%, even when all the 
measured Higgs couplings with gauge bosons and fermions are 
consistent with the SM values. 
At LC's, such a large difference in the $hhh$ coupling 
from the SM prediction may be detected.  
In the weakly coupled THDM, the one-loop effect is small 
and decouples in the large mass limit for the heavier 
Higgs bosons.
The quartic mass effect on the effective 
$hhh$ coupling is a general characteristic in any new physics 
model which has the non-decoupling property.

\vspace*{4mm}
\noindent
{\bf ACKNOWLEDGMENTS}

\vspace*{2mm}
\indent
The work of Y.O. was supported by Grant-in-Aid for 
Scientific Research (C)(No. 13640309).
The work of C.P.Y. was supported by the NSF Grant PHY-0100677.


\end{document}